\documentclass[fleqn,usenatbib]{mnras}

\usepackage{newtxtext,newtxmath}
\usepackage[T1]{fontenc}

\DeclareRobustCommand{\VAN}[3]{#2}
\let\VANthebibliography\thebibliography
\def\thebibliography{\DeclareRobustCommand{\VAN}[3]{##3}\VANthebibliography}

\usepackage{graphicx}	% Including figure files
\usepackage{amsmath}	% Advanced maths commands
\usepackage{pdflscape}  % Make landscape table possible
\usepackage{subcaption}

\title[Host galaxies of two TDE candidates]{Host Galaxy Line Diagnostics for the Candidate Tidal Disruption Events XMMSL1~J111527.3$+$180638 and PTF09axc}

\author[A. Inkenhaag et al.]{
Anne~Inkenhaag,$^{1,2}$\thanks{E-mail: a.inkenhaag@astro.ru.nl}
Peter~G.~Jonker,$^{1,2}$
Giacomo~Cannizzaro,$^{1,2}$
Daniel~Mata~S\'anchez,$^{3}$
Richard~D.~Saxton$^{4}$
\\
$^{1}$Department of Astrophysics/IMAPP, Radboud University Nijmegen, P.O.~Box 9010, 6500 GL Nijmegen, The Netherlands\\
$^{2}$SRON, Netherlands Institute for Space Research, Sorbonnelaan 2, 3584~CA, Utrecht, The Netherlands\\
$^{3}$Jodrell Bank Centre for Astrophysics, Department of Physics and Astronomy, The University of Manchester, M13 9PL, UK\\
$^{4}$Telespazio for ESA, XMM-Newton SOC, ESAC, Apartado 78, 28691 Villanueva de la Ca\~nada, Madrid, Spain
}

\date{Accepted XXX. Received YYY; in original form ZZZ}

\pubyear{2021}

\begin{document}
\label{firstpage}
\pagerange{\pageref{firstpage}--\pageref{lastpage}}
\maketitle

\begin{abstract}
We present results of our analysis of spectra of the host galaxies of the candidate Tidal Disruption Events (TDEs) XMMSL1~J111527.3$+$180638 and PTF09axc to determine the nature of these transients. We subtract the starlight component from the host galaxy spectra to determine the origin of the nuclear emission lines. Using a Baldwin--Phillips--Terlevich (BPT) diagram we conclude that the host galaxy of XMMSL1~J111527.3$+$180638 is classified as a Seyfert galaxy, suggesting this transient is likely to be caused by (extreme) variability in the active galactic nucleus. We find that the host of PTF09axc falls in the 'star-forming' region of the BPT-diagram, implying that the transient is a strong TDE candidate. For both galaxies we find a \textit{WISE}-colour difference of $W1-W2<0.8$, which means there is no indication of a dusty torus and therefore an active galactic nucleus, seemingly contradicting our BPT finding for the host of XMMSL1~J111527.3$+$180638. We discuss possible reasons for the discrepant results obtained through the two methods. 
\end{abstract}

\begin{keywords}
transients: tidal disruption events -- galaxies: individual: NGC~3599 -- galaxies: active -- galaxies: nuclei -- black hole physics
\end{keywords}

%%%%%%%%%%%%%%%%%%%%%%%%%%%%%%%%%%%%%%%%%%%%%%%%%%

%%%%%%%%%%%%%%%%% BODY OF PAPER %%%%%%%%%%%%%%%%%%

\section{Introduction} \label{intro}

%The process of tidal disruption of a star was first studied as a fuel mechanism for active galactic nuclei (AGNs) by \citet{Hills1975}. \citet{Rees1988} (and \citet{Phinney1989} with the correct accretion fall back time-scale) suggested these tidal disruption events (TDEs) could be an identifier for the presence of quiescent SMBHs in other galaxies, which would be hard to observe otherwise. These events have the distinctive signature of an accretion powered flare lasting up to a few years. 

%They are events where a star moves too close to a supermassive black hole (SMBH), causing the gravity of the SMBH to tidally disrupt the star. 
Two-body relaxation processes in the nucleus of a galaxy make stars wander in energy and momentum space. This can bring the pericentre of a star's orbit within its tidal radius (or Roche limit) given the supermassive black hole (SMBH) in the centre of the galaxy, causing the difference in gravitational pull between the part of the star nearest and furthest from the SMBH, also known as the tidal force, to overcome the self-gravity of the star. This results in the star being pulled apart in a tidal disruption event (TDE) \citep{Hills1975, Rees1988, Phinney1989}. 

Part of the stellar material of the disrupted star will stay bound to the SMBH and accrete on to it, creating a luminous flare that is visible across the electromagnetic spectrum \citep{Rees1988, Lodato2011}. Over the last two decades, dozens of TDEs have been classified from among transient nuclear flares detected in X-ray, optical or UV (see \citealt{VanVelzen2020, Saxton2021} for a review). Optically detected TDEs often go undetected in X-rays \citep{Gezari2012}, and vice versa. Although, there are events that have been found to emit in optical and X-rays (e.g., ASASSN-14li, ASASSN-15oi, AT~2019dsg, AT~2018fyk; \citealt{Holoien2016, Gezari2017, Cannizzaro2020b, Wevers2019}, respectively).% and some events have been found to emit both X-rays and optical light, albeit not simultaneously \citep{Jonker2020}. 

A list of properties need to be satisfied for both optical/UV and X-ray selected TDE candidates to be confirmed. This list is based on observational characteristics shared by the known population of TDEs, and it is refined over time (see \citealt{Zabludoff2021} for a review). Key observables for optical/UV TDEs include broad He and/or H lines and blue continuum emission (see \citealt{VanVelzen2020} for a review). A short rise to peak, steady decline and a soft X-ray spectrum are among the key observables for X-ray selected TDEs (see \citealt{Saxton2021} for a review). There are multiple competing models explaining the optical-UV emission mechanism: outflows (photon-driven, \citealt{Strubbe2009}, line-driven, \citealt{Miller2015} or circularisation-driven, \citealt{Metzger2016}), reprocessing of accretion disc emission by material in the debris stream at larger radii \citep{Guillochon2014} or shocks in the self-intersecting debris stream \citep{Piran2015, Bonnerot2017}. The lack of agreement about the importance of, for instance, the self-intersection shock, the rate of circularisation of the stellar debris, and the accretion radiation efficiency makes that there is no single theoretical prediction that can serve as a guideline to classify an event as a TDE. While some of the observed properties can be explained by the theoretical models under consideration, there might be TDEs that do not fit in the sample of previously classified TDEs, for instance if they occupy a different part of the parameter space such as penetration factor ($\beta$), SMBH mass or spin, or stellar mass. Therefore, we need to keep a critical but open mind about which transients we classify as TDEs.

%For both optical and X-ray discovered TDE candidates, there is a list of properties that need to be satisfied for a transient to be classified as a TDE. These properties are observational characteristics among the sample of events that have already been classified as TDEs and the list of characteristics has grown in time from the observed events.
 
TDEs are often detected in otherwise inactive galaxies. However, \citet{Kennedy2016} suggest the TDE rate in AGNs could be enhanced with a factor up to 10 due to the interaction of stars with the disc around the AGN.  Detecting TDEs in galaxies hosting an active galactic nucleus (AGN) is difficult, due to the inherent difficulty in distinguishing them from regular AGN activity, although some TDEs have been discovered in low-luminosity AGNs (e.g., ASASSN-14li and AT2019qiz; \citealt{Holoien2016, Nicholl2020}, respectively) and even higher luminosity AGNs (e.g., PS16dtm and SDSS~J015957.64$+$003310.5; \citealt{Blanchard2017, Merloni2015}, respectively). When confronted with AGNs at the same redshift, TDEs are typically brighter \citep{Auchettl2018}, but more extreme AGN variability can be as luminous as a TDE flare \citep[e.g.,][]{Cannizzaro2020a}. This emphasizes the difficulty in distinguishing between TDEs and AGN flares. Besides this, the interaction between the TDE debris stream and the AGN disc and the effect on the emitted luminosity are currently not well understood (although see \citealt{Chan2019, Chan2020} for modelling). Current theoretical models of the interaction of the stream originating in the destruction of a star and a pre-existing AGN disc are uncertain as they sample a restricted section of the parameter space and do not run long enough to study the accretion of an important fraction of the TDE debris. Finding a TDE candidate in a quiescent galaxy means there is one less alternative explanation for the transient.

%One of the properties of the observed sample of both the optical-UV selected and the X-ray selected TDEs is that the host galaxy appears to be quiescent and inactive before and after the transients, so not hosting an active galactic nucleus (AGN). Recently, TDEs have been reported to happen in low-luminosity AGNs (e.g., ASASSN-14li \citep{Holoien2016} and AT2019qiz \citep{Nicholl2020}). These TDEs can be distinguished from AGN flares, as the AGNs in their host galaxies have a low luminosity. %Classifying and studying TDEs in other AGNs is hindered by the difficulty in distinguishing between TDEs and AGN flares, which could explain the low number of confirmed TDEs in galaxies hosting an AGN (e.g., PS16dtm and SDSS~J015957.64$+$003310.5 \citealt{Blanchard2017, Merloni2015}, respectively). 
%\citet{Auchettl2018} investigate the similarities and differences in X-ray emission between TDEs and AGN flares. Among other characteristics, they find that at their peak luminosity, TDEs are brighter than AGNs at the same redshift. Highly variable AGNs on the other hand are able to produce flares of the same order of magnitude as those seen from X-ray TDEs, see e.g., \citet{Cannizzaro2020a}. 

A galaxy is classified as an AGN/non-quiescent galaxy if one or more of the following properties is observed \citep{VanVelzen2020}: \textit{i)} The luminosity from the nucleus of the galaxy varied significantly with time before the main flare/transient event e.g., in the optical or X-ray luminosity, \textit{ii)} The \textit{WISE}-colours indicate the presence of a dusty torus, $W1-W2~\geq~0.8$ \citep{Stern2012}, \textit{iii)} The ratio between the equivalent widths (EWs) of specific emission lines in the optical -- restframe -- part of the nuclear host spectrum show that the source falls in the AGN region of the Baldwin--Phillips--Terlevich (BPT) diagram. The ratios of EWs of emission lines reflect the physical conditions under which these lines were formed. These conditions are different for the different options considered for their formations (e.g., AGN, star-forming regions, LINER-like shocks) \citep{BPT1981, CidFernandes2009}. The AGN-region in the BPT-diagram is the region where the ionization-mechanism is dominated by the -- UV / X-ray -- ionizing radiation from the AGN \citep{BPT1981}.  

%the diagnostic power of the ratios between the equivalent widths of various emission lines present in the (restframe) optical part of the nuclear host spectrum, because those reflect the physical conditions under which these lines are formed. These conditions are different for the different options considered for their formations (e.g., AGN, star forming regions, LINER-like shocks) \citep{BPT1981, CidFernandes2009}.

In this work we investigate the host galaxy of two TDE candidates; the X-ray discovered TDE candidate XMMSL1~J111527.3$+$180638 in the galaxy NGC~3599 and the optically discovered candidate PTF09axc in the galaxy SDSS~J145313.07$+$221432.2. The nature of the observed flare -- AGN activity or TDE -- has been subject of discussion in the literature for both of these candidates (e.g., \citealt{Saxton2015} for XMMSL1~J111527.3$+$180638 and \citealt{Arcavi2014,Jonker2020} for PTF09axc). %The goal of this paper is to get a clearer view of the nature of these two candidate TDEs and, if possible, to draw a conclusion on whether these two candidates are actual TDEs or a different type of transient.

We aim to classify the host galaxies of two TDE candidates by determining the position of the nuclear emission region on BPT-diagrams. We also look at the \textit{WISE}-colours of the host galaxies and we use the existing $L_X~\propto$~[\ion{O}{iii}] correlation observed in AGNs \citep{Heckman2005} to compare the observed [\ion{O}{iii}]~$\lambda$5007 luminosity to what is predicted on the basis of the correlation. We finally compare the luminosity expressed in units of the Eddington luminosity of NGC~3599 with that of other low-luminosity AGN host galaxies of TDEs.

%the observed [\ion{O}{iii}]~$\lambda$5007 and X-ray luminosity with those predicted on the basis of the existing $L_X~\propto$~[\ion{O}{iii}] correlation observed in AGNs \citep{Heckman2005}.

%To achieve our goal we use the ratios of various emission lines in the optical part of the spectrum of the nuclear region of the host galaxies to investigate using a BPT diagram the nature of the line ionization mechanism. We also look at the WISE-colours of the host galaxies and for the host of PTF09axc we compare the observed [\ion{O}{iii}]~$\lambda$5007 and X-ray luminosity with those predicted on the basis of the existing $L_X~\propto$~[\ion{O}{iii}] correlation observed in AGNs \citep{Heckman2005}.

\section{Data}\label{Data}

\subsection{XMMSL1~J111527.3+180638} \label{DataNGC3599}

XMMSL1~J111527.3$+$180638 -- XMMJ1115 from now on -- was first reported as a candidate TDE based on an X-ray flare seen in \textit{XMM-Newton} slew data on 2003 November 22 \citep{Esquej2007}. Its associated host galaxy is NGC~3599, at redshift $z=0.0028$, $d_\text{L} = 19.86$~Mpc taken from the Cosmicflows-3 Distance Catalogue \citep{Tully2016}. Flux calibrated optical spectra were obtained from \citet{Esquej2008}, originally \citet{Caldwell2003}. This data set consists of two spectra: a blue spectrum with wavelengths 3500--5500~$\text{\AA}$ taken on 1998 May 20 and a red spectrum with wavelengths 5500--7500~$\text{\AA}$ taken on 2000 February 5, both taken at the F.L.~Whipple Observatory (Mount Hopkins near Amado, Arizona, USA) with the FAST (FAst Spectrograph for the Tillinghast Telescope) instrument located on the 1.5-m Tillinghast telescope. Both spectra were taken before the reported X-ray flare (see \citealt{Caldwell2003} for full observational details and the data reduction procedure).

\subsection{PTF09axc}

PTF09axc was first reported by \citet{Arcavi2014} as part of their archival search of the Palomar Transient Factory (PTF) data for blue transients with $-21 \leq M_{R (peak)} \leq -19$. The discovery date is 2009 June 20 and the source is associated with the galaxy SDSS~J145313.07$+$221432.2 -- hereafter SDSSJ1453 -- at redshift $z=0.115$ \citep{Arcavi2010}. 

We have taken two 1800~s low resolution optical spectra of the nucleus of the host galaxy after the transient had faded \citep{Arcavi2014} on 2019 July 14 and 15 using the Auxiliary-port CAMera (ACAM) mounted at the Cassegrain focus of the William Herschel Telescope (WHT) located at the Roque de los Muchachos Observatory on La Palma, Spain under program W19AN009. Using the V400 grating, GG395A order blocking filter, and the AUXCAM CCD results in a wavelength coverage of $3950$--$9400$~$\text{\AA}$ and resolution R$\sim430$ for a 1~arcsec slit. We correct for instrumental broadening during the analysis of these data. %change to \prime\prime for MNRAS% 

Data reduction is done using a program written in {\sc python} that uses {\sc lacosmic} \citep{vanDokkum2001} for cosmic ray cleansing, {\sc pyraf} for bias and flatfield corrections and {\sc molly}, developed by T.~Marsh \citep{Marsh2019}\footnote{http://deneb.astro.warwick.ac.uk/phsaap/software/molly/html/INDEX.html}, for wavelength calibrations. We further use {\sc molly} to flux calibrate and average our spectra. 

\section{Analysis and Results}\label{Results}

\begin{figure*}
 \centering
 \hspace*{-.7cm}\includegraphics[width=0.52\textwidth]{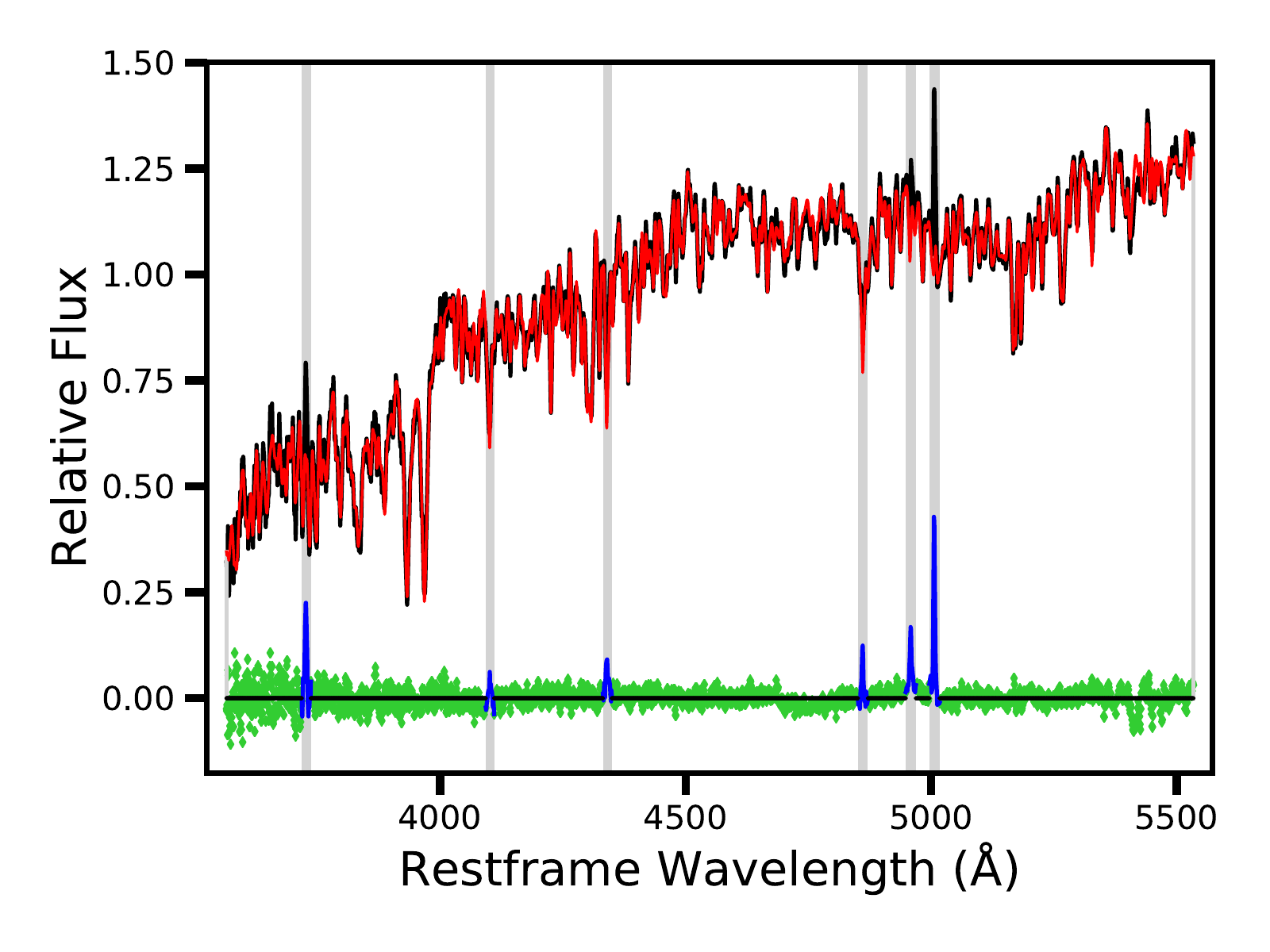}
 \includegraphics[width=0.50\textwidth]{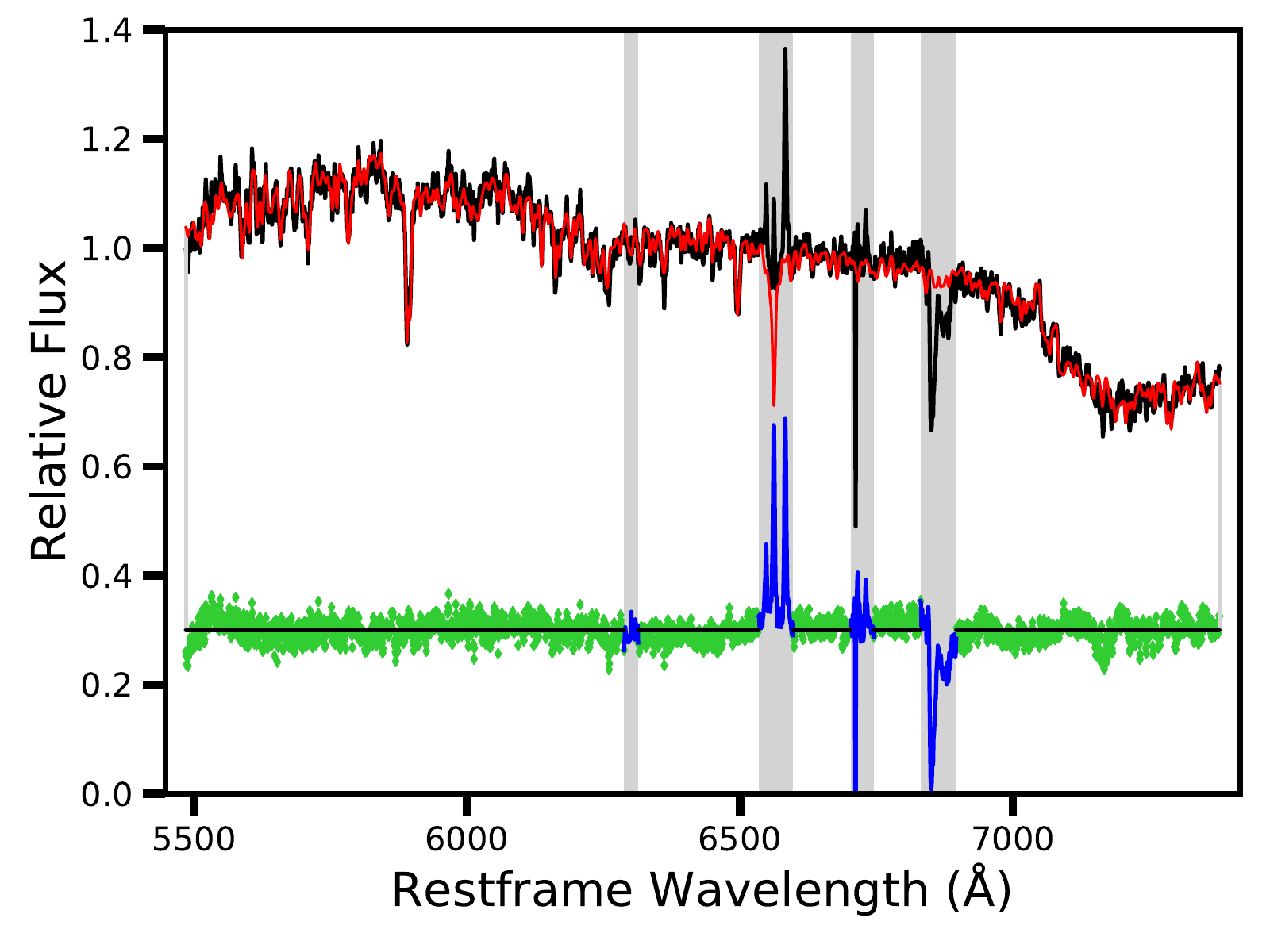}
  \caption{The normalised spectrum of NGC~3599 -- host of XMMJ1115 -- is shown in black, the best fit for the starlight component from {\sc ppxf} is over-plotted in red, and the residuals after subtracting the best-fitting host-galaxy stellar spectrum is shown in green and blue. The grey bands -- corresponding to the blue wavelength regions in the subtracted spectrum -- are wavelengths around relevant host galaxy emission lines or telluric absorption lines expected from the night sky and they are therefore excluded from the {\sc ppxf} fit. \textit{Left panel:} The blue part of the spectrum (3500--5500~$\text{\AA}$) displaying the masked emission lined from H~$\beta$~$\lambda$4861 and the [\ion{O}{iii}]~$\lambda4959,~5007$ doublet in grey. \textit{Right panel:} The red part of the spectrum (5500--7500~$\text{\AA}$) displaying the masked emission lines from H~$\alpha$~$\lambda$6563, the [\ion{N}{ii}]~$\lambda$6548,~6584 doublet and the [\ion{S}{ii}]~$\lambda$6617,~6631 doublet in grey. The region around [\ion{O}{i}]~$\lambda6300$ is also masked, but there is no emission line detected when trying to fit a Gaussian at this wavelength.}
 \label{pPXF-NGC3599}
\end{figure*}

To obtain an accurate nuclear source classification we start by subtracting the starlight component from the host galaxy spectrum using the Penalized PiXel-Fitting ({\sc ppxf}) method \citep{Cappellari2017}, used with the MILES stellar library \citep{Vazdekis2010}. We use a degree four multiplicative Legendre polynomial -- as opposed to an additive polynomial -- to correct the continuum shape during the fit to prevent changes in the line strength of the absorption features in the templates, to minimize the influence on the strength of any emission line in the nuclear spectrum after subtraction. After subtraction the continuum emission is reduced to zero which means the equivalent width -- flux in a line divided by the continuum -- becomes undefined. Instead, we use the flux of the emission lines to determine the source position in a BPT-diagram. We use the python package {\sc lmfit} to fit Gaussian curves to the stellar-host subtracted emission -- or absorption -- lines to obtain the flux of the following emission lines of interest for the BPT-diagram(s), where present: H~$\alpha$~$\lambda$6563, H~$\beta$~$\lambda$4861, [\ion{O}{iii}]~$\lambda4959,~5007$, [\ion{O}{i}]~$\lambda6300$, [\ion{N}{ii}]~$\lambda$6548,~6584, [\ion{S}{ii}]~$\lambda$6617,~6631. To reduce the number of degrees of freedom during fitting we require the Full Width at Half Maximum (FWHM) of lines in doublets to be the same. We also fix the wavelength separation of doublets to their laboratory value and we fix the ratio in amplitudes for the lines in doublets when an amplitude ratio in known ([\ion{O}{iii}]~$\lambda5007/$[\ion{O}{iii}]~$\lambda4959=3$ and [\ion{N}{ii}]~$\lambda6584/$[\ion{N}{ii}]~$\lambda6548=3$ from \citealt{Osterbrock2006}). % In this process we tie the Full Width at Half Maximum (FWHM) of the narrow lines close in wavenlength to be the same (after correcting for instrumental broadening when necessary), so we tie the FWHM of narrow component of H~$\alpha$ and the [\ion{N}{ii}] doublet together, the narrow component of H~$\beta$ and the [\ion{O}{iii}] doublet together and the [\ion{S}{ii}] doublet. We expect the FWHM of all narrow emission line to be the same, within errobars, if this is not the case we use the FWHM of H~$\alpha$ and the [\ion{N}{ii}] doublet as the FWHM of the narrow line during the fit, as these are the most prominent emission lines in the spectra.  Furthermore we tie the amplitudes of the lines of the doublets together according to \textbf{REFERNCE WITH LINE RATIONS FOR N2, O3 and S2}. The separation between the lines in the doublets have been fixed during the fitting procedure. 

\subsection{XMMJ1115}
\label{analysis-NGC3599}

Fig.~\ref{pPXF-NGC3599} shows the normalised -- divided by the median value -- galaxy spectrum in black, with the best fit from {\sc ppxf} over-plotted in red, the grey bands represent areas masked during the fitting procedure. The red and blue parts, as described in Section~\ref{DataNGC3599}, of the spectrum are fitted separately. The best fitting stellar population has a redshifted radial velocity of $799\pm2$~km~s$^{-1}$ (we average the radial velocity derived from the red and the blue parts of the spectrum). Taking into account the average rms uncertainty of 58~km~s$^{-1}$ on the wavelength calibration (from \citealt{Caldwell2003}) the redshift we derive is in agreement with previous measurements for this galaxy (e.g., $839\pm5$~km~$^{-1}$; \citealt{Cappellari2011}). Subtracting the starlight component leaves us with the nuclear emission line spectrum (blue/green in Fig.~\ref{pPXF-NGC3599}). 

We detect emission lines of H~$\alpha$, H~$\beta$ and the [\ion{O}{iii}], [\ion{N}{ii}] and [\ion{S}{ii}] doublets, but we do not detect a significant emission line for [\ion{O}{i}] in this source. The detected emission lines are best fitted with one Gaussian component with an average FWHM of $253\pm9$~km~s$^{-1}$ (see Table \ref{tab:lines} in the appendix). We provide figures showing the Gaussian fits to the detected emission lines in Fig.~\ref{NGC3599_lines} in the appendix.

\begin{figure}
 \includegraphics[width=\columnwidth]{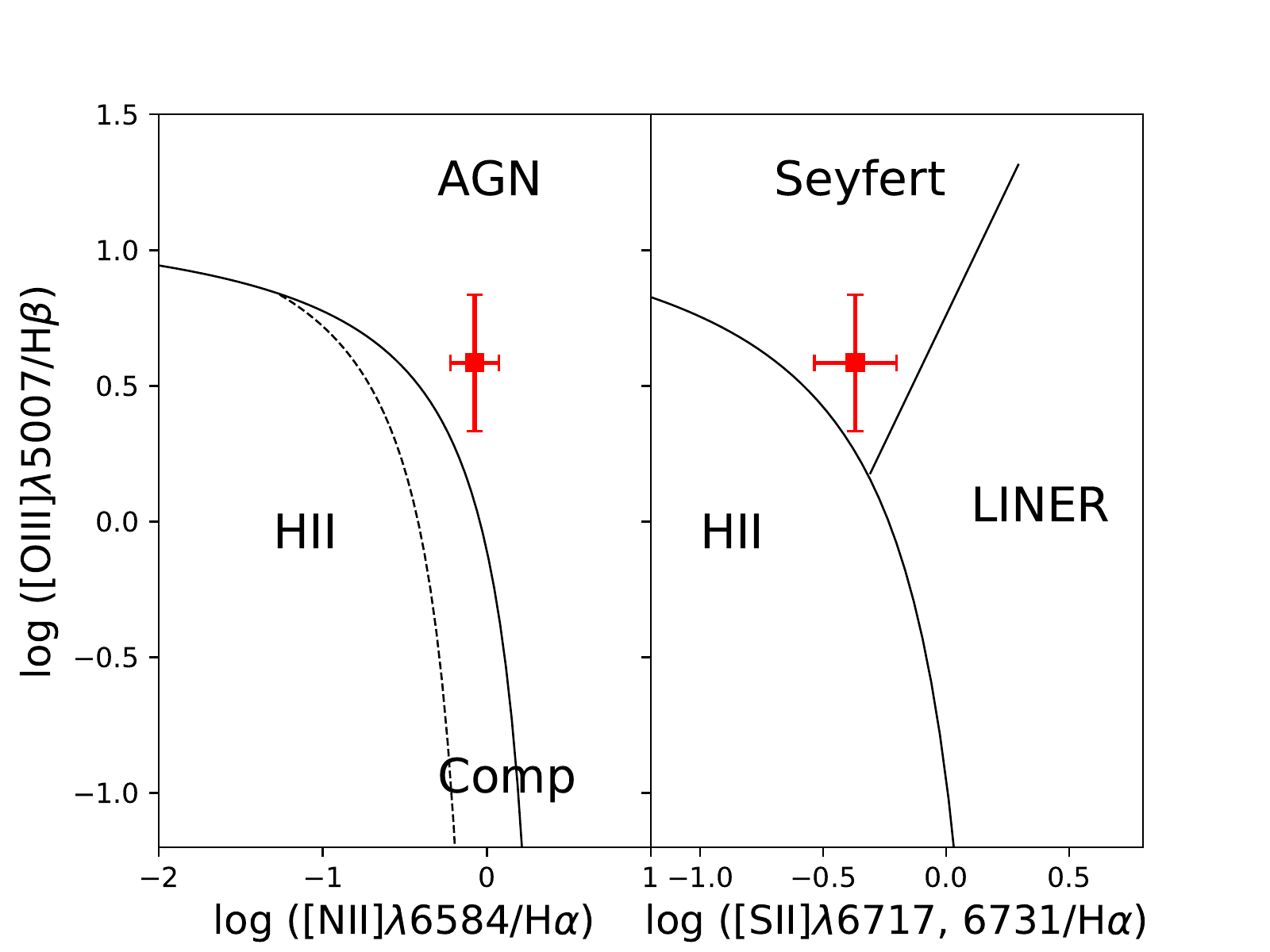}
 \caption{Baldwin--Phillips--Terlevich (BPT) diagram \citep{BPT1981} for the nucleus of NGC~3599 derived using the flux measurements of the emission lines detected in the pre-outburst spectrum. The left panel uses the ratio between the [\ion{O}{iii}]~$\lambda5007$ and the  H~$\beta$ emission line flux and the ratio between the [\ion{N}{ii}]~$\lambda6584$ and the H~$\alpha$  emission line flux while the left panel used the ratio between the [\ion{O}{iii}]~$\lambda5007$ and H~$\beta$  emission line flux and the ratio between the [\ion{S}{ii}]~$\lambda6617,~6731$ and H~$\alpha$  emission line flux. The demarcations between different regions caused by different ionization mechanisms are from \citet{Kewley2001, Kauffmann2003, Kewley2006}. The position of the source in both diagrams is consistent with an AGN/Seyfert galaxy being present in NGC~3599 prior to the flare.}
 \label{BPT-NGC3599}
\end{figure}

Using the flux of the emission lines we calculate the position of the source in a BPT-diagram. We use the demarcations from \citet{Kewley2001, Kauffmann2003, Kewley2006} to indicate different ionization-mechanism regions. The location of NGC3599 in the BPT-diagram is shown in Fig.~\ref{BPT-NGC3599} and it is consistent with an AGN/Seyfert classification for the spectrum of the nuclear region of the host of XMMJ1115.  

\subsection{PTF09axc}

\begin{figure}
 \hspace*{-.7cm}\includegraphics[width=1.1\columnwidth]{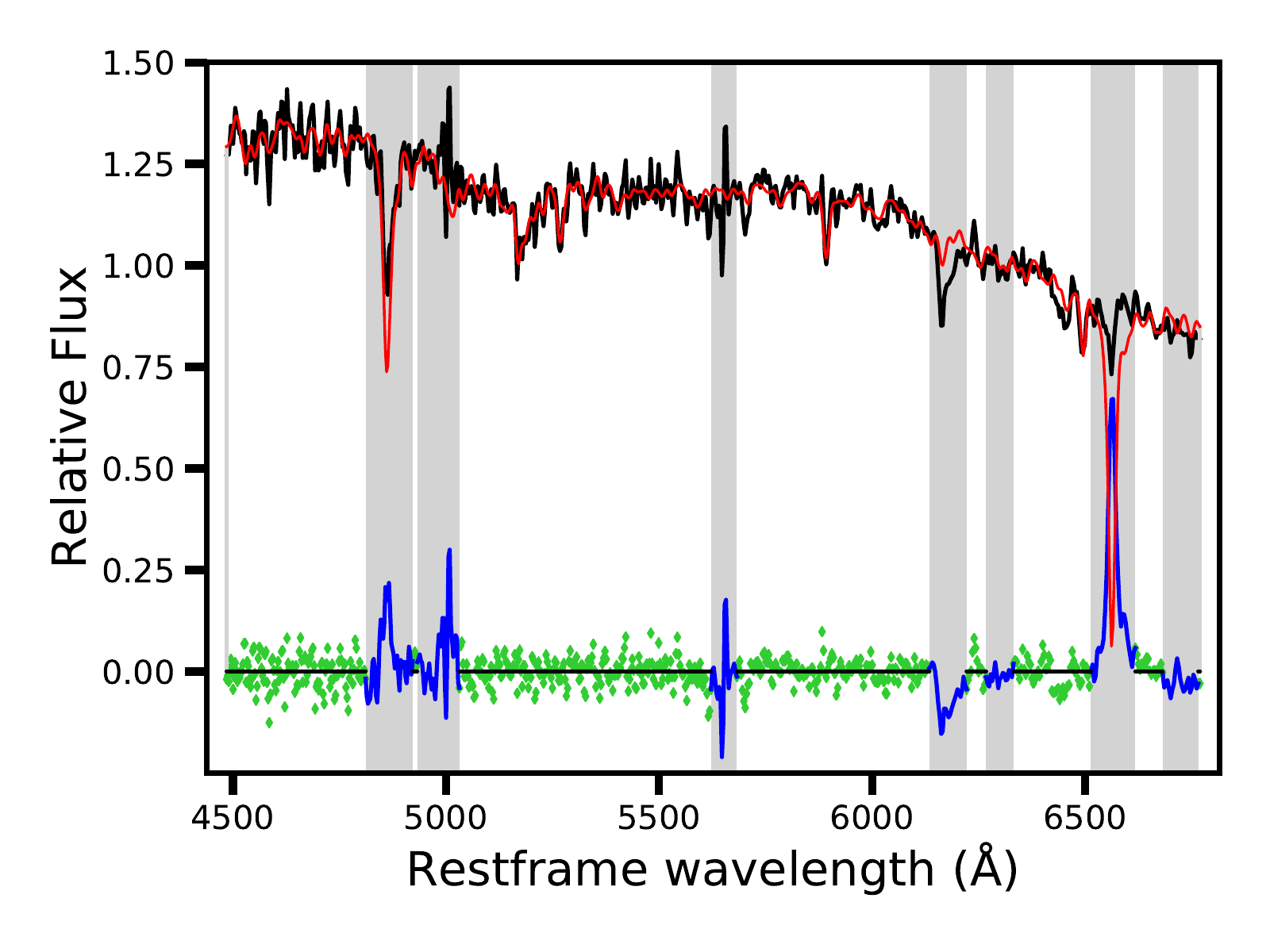}
 \caption{The normalised spectrum of SDSSJ1453 -- host of PTF09axc -- with the same colour scheme as in Fig.~\ref{pPXF-NGC3599}. The masked emission lines in grey are H~$\alpha$~$\lambda$6563, H~$\beta$~$\lambda$4861, the [\ion{O}{iii}]~$\lambda4959,~5007$ doublet and the [\ion{N}{ii}]~$\lambda$6548,~6584 doublet. {\sc pPXF} also masks [\ion{O}{i}]~$\lambda6300$ and  [\ion{S}{ii}]~$\lambda$6617,~6631, but there are no visible emission lines present in the subtracted spectrum at those wavelengths.}
 \label{pPXF-PTF09axc}
\end{figure}

We repeat the exact same data analysis procedure we employed for NGC~3599 for the host of PTF09axc (SDSSJ1453), see Section~\ref{analysis-NGC3599}. Blueshifting the spectrum with $z=0.1153$ before using {\sc ppxf} leaves a residual radial velocity of $9\pm21$~km~s$^{-1}$, which is consistent with no residual radial velocity. Therefore, the best-fitting stellar population has redshift of $z=0.1153\pm0.0001$, corresponding to radial velocity $32590\pm30$~km~s$^{-1}$. For an uncertainty in the last digit of the redshift given in \citet{Arcavi2010} (confirmed through private communication) as small as 1, our values are completely consistent with their redshift. This redshift corresponds to $d_\text{L} = 536.1$~Mpc using $\Omega_m=0.3$, $\Omega_\Lambda= 0.7$ and H$_0 = 70$~km~s$^{-1}$~Mpc$^{-1}$

We do not detect emission lines for [\ion{O}{i}] and the [\ion{S}{ii}] doublet, but we do detect the [\ion{N}{ii}] doublet and the H~$\alpha$ and H~$\beta$ emission lines. To ensure the Gaussian functions fit to H~$\alpha$ and the [\ion{N}{ii}] doublet have the correct centra wavelength, we have to fix the wavelength separation between the lines. We fit a single component with an (average) FWHM=$768\pm102$~km~s$^{-1}$ to the detected emission lines. We derive an upper limit to the flux for both lines in the [\ion{O}{iii}] doublet. 

%There is an average shift in the central wavelength of the detected lines of $3.9~\textrm{\AA}$, which is likely due to rms uncertainty in the wavelength calibration of the data. 

%(taking the average of the shift in the detected lines of $3.9~\textrm{\AA}$)

%for [\ion{O}{iii}]~$\lambda5007$ while [\ion{O}{iii}]~$\lambda4959$ is not detected.

%As the Balmer decrement is typically larger than 1 \textbf{REFERENCE}, and given the higher sensitivity of our spectrum around the wavelengths of H~$\alpha$ than around H~$\beta$, the lack of detection of a broad component to the H~$\alpha$ emission line makes the detection of a broad component to the H~$\beta$ line improbable. This means we will have to force an emission line to fit which has a smaller FWHM then would fit if we leave this parameter free in the fit. We require the FWHM of the H~$\beta$ line to have a value within the range FHMW$\pm~1~\sigma$ given by the Gaussian fits to the other narrow lines. From te narrow lines we select the FWHM and 1~$\sigma$ from the one with the best fit, i.e., the one with the lowest reduced $\chi^2$. \textbf{Add why we choose this one. }

Due to the redshift of this source the [\ion{O}{iii}]~$\lambda5007$ emission line is redshifted to $\sim5584$~$\textrm{\AA}$ which is close to the wavelength of the [\ion{O}{i}]~$\lambda5577$ terrestrial sky emission line. In fact, these two lines fall within one ACAM resolution element of each other. In order to obtain as accurate as possible an upper limit on the presence of the [\ion{O}{iii}]~$\lambda5007$, we tried several data reduction optimisations tailored to allow as clean a subtraction of the terrestrial sky emission line as possible. To make sure the sky lines are perpendicular to the spectral trace we extracted a rectified version of the 2-D spectrum. This did not significantly improve the subtraction of the sky emission line. Next, we used the {\sc fit2d} option during the {\sc iraf} {\sc apall} procedure to extract the spectrum, which uses a two dimensional function to smooth the profile to use with variance weighting or cleaning. However, this also did not significantly improve the subtraction of the sky emission line either. Therefore, we proceed with the spectrum obtained from the original data reduction process, with the one difference that for determining the upper limit to the [\ion{O}{iii}] doublet emission line we do not subtract the sky emission lines. 

\begin{figure}
 \includegraphics[width=\columnwidth]{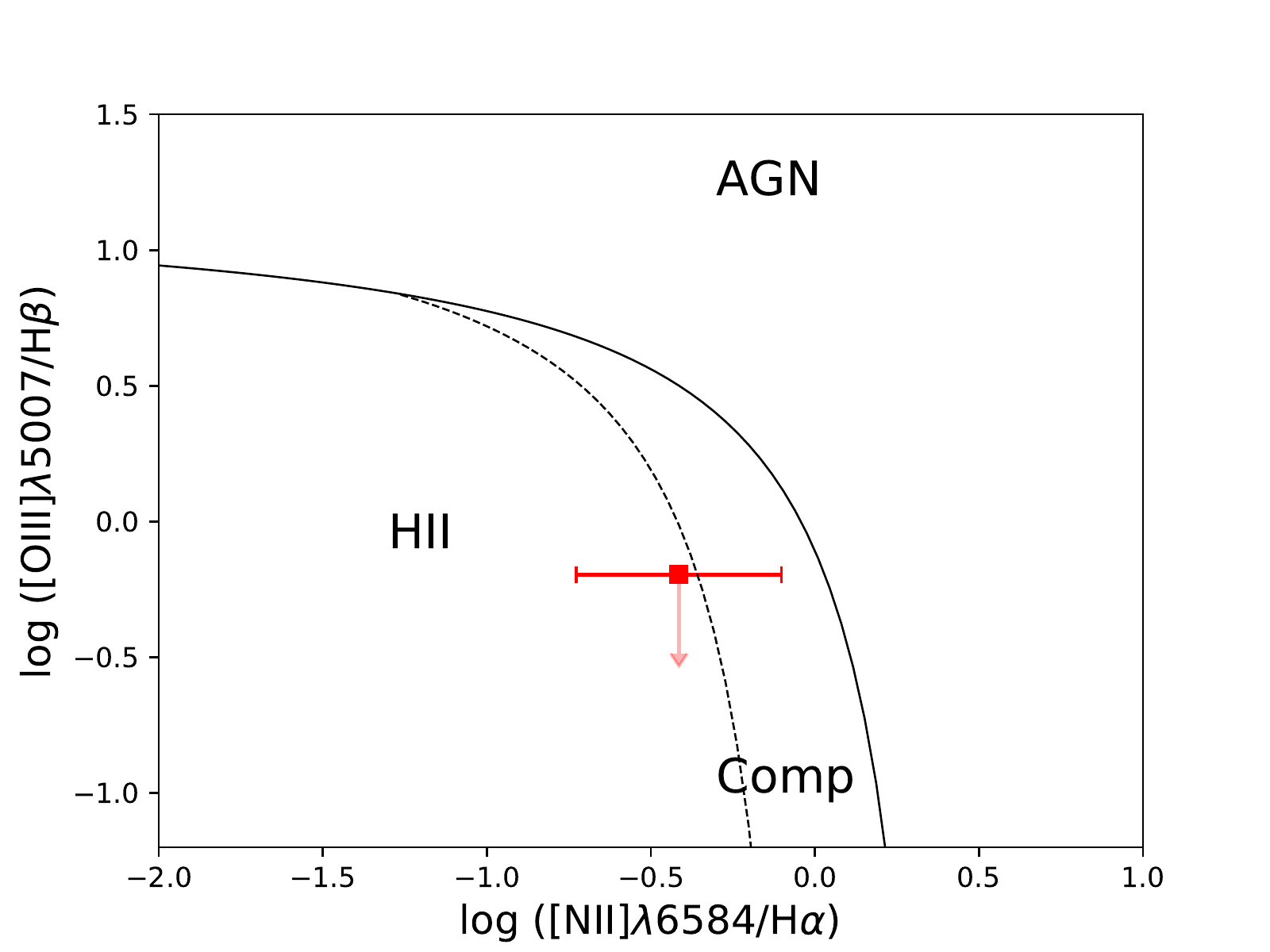}
 \caption{Baldwin--Phillips--Terlevich (BPT) diagram \citep{BPT1981} using the flux measurements for the emission lines detected in the post-outburst spectrum of SDSSJ1453. It uses the ratio between the [\ion{O}{iii}]~$\lambda5007$ and H~$\beta$ emission line flux and the ratio between the  [\ion{N}{ii}]~$\lambda6584$ and H~$\alpha$ emission line flux. The position of the source is consistent with an H~II/star forming region or a composite region within the error bars.}
 \label{BPT-PTF09axc}
\end{figure}

Instead, we derive an upper limit for the flux in the [\ion{O}{iii}]~$\lambda5007$ line by fitting two Gaussians, one to the $5577$~$\textrm{\AA}$ sky line and one to the [\ion{O}{iii}]~$\lambda$5007 line. We shift the spectrum back to the rest frame wavelength of the host galaxy, this includes the $5577$~$\textrm{\AA}$ sky line. We fix the central wavelength of the Gaussian function designed to describe this sky line to the expected value after blueshifting this line to the galaxy rest frame. In addition, we fix its FWHM to the spectral resolution of ACAM, leaving only the amplitude as a free parameter for this Gaussian during the fit. For the Gaussian designed to determine the upper limit to the [\ion{O}{iii}]~$\lambda5007$ line, we fix the central wavelength to where we expect it to appear. 
%including the small additional shift from its rest frame wavelength as found for the other narrow emission lines (taking the average of the shift in the detected lines of $3.9~\textrm{\AA}$).
We fix the FWHM of this emission line to the average value measured in the other lines. This leaves only the amplitude as a free parameter in the fit for this Gaussian. We fit the [\ion{O}{iii}]~$\lambda$4959 line simultaneously, with the same FWHM as [\ion{O}{iii}]~$\lambda$5007 and the wavelength separation between the lines of the doublet set to the laboratory value, leaving only the amplitude free during the fit. 
%To obtain the upper limits we take the flux in the emission line plus two times the error in the flux (if this end product is negative we take the absolute value of this end value).
As the upper limit on the flux of the Gaussian-shaped emission line is determined from a one sided Gaussian probability distribution, the 2$\sigma$ upper limit corresponds to the 95 per cent confidence level. We use this upper limit to derive the position of the source in a BPT-diagram (see Fig.~\ref{BPT-PTF09axc}). The source falls in the region of the BPT-diagram associated with star-forming and H~II galaxies.
%on the line separating the star-forming/H~II galaxies and the composite galaxies and is consistent with either classification within 1$\sigma$. 
% %

\section{Discussion}\label{Discussion}

In this paper we investigate the nature of two candidate TDE events, XMMJ1115 and PTF09axc, by classifying the nuclear regions of their host galaxies, NGC~3599 and SDSSJ1453, respectively. We apply two methods: optical emission lines ratios to assess the ionization-mechanism and the infrared (IR) colours as determined by the \textit{WISE}-satellite \citep{WISE2010} to investigate if dust, as often found in a dusty torus in an AGN, is present. The dusty torus of an AGN will yield $W1-W2 \geq 0.8$ \citep{Stern2012}, where $W1$ and $W2$ indicate the \textit{WISE}-bands at 3.4 and 4.6~$\mu$m, respectively. For SDSSJ1453, we apply a third method using the empirical relation between $L_\text{X}$(3--20~keV) and $L_{\text{[\ion{O}{iii}}]}$ from \citet{Heckman2005} to assess if the observed X-ray luminosity is consistent with that from an AGN assuming the [\ion{O}{iii}] emission is caused by the AGNs narrow line region. 

%For NGC~3599 we also use this relation, but for this source we use it to calculate the X-ray luminosity expressed as a fraction of the Eddington luminosity from the observed $L_{\text{[\ion{O}{iii}}]}$, assuming NGC~3599 is an AGN. Next, we compare it to the Eddingtion ratio of low-luminosity AGNs that are confirmed TDE hosts. %similar to the environment of confirmed TDEs in low-luminosity AGNs. 

The position of NGC~3599 on the BPT-diagram suggests it is a Seyfert galaxy, and there have been more papers suggesting the host galaxy is not a quiescent galaxy, see e.g., \citet{Saxton2015}. \citet{Saxton2015} show that the galaxy was luminous in X-rays 18 months before the peak flux was measured, showing it to be bright on much longer timescales than shown by TDEs known at that point in time. They also argue that even if one of the two measurements was taken during rise-time and one during decay, the rise-time and the plateau phase together would still be significantly longer than seen in previous TDE candidates at that time. Since then, however, longer lived TDE-candidates have been observed (e.g., \citealt{Lin2017}) which means the measurements by \citet{Saxton2015} can no longer be considered unusual behaviour for a TDE. With our current understanding of X-ray TDE light curves we can therefore not make a definitive distinction between a TDE or AGN-activity for the flare XMMJ1115. However, our work does strengthen the evidence that the nucleus of NGC~3599 hosts a low-luminosity AGN.

%We therefore prefer AGN-variability as an explanation for the observed flare.

We calculate $W1-W2 = -0.032 \pm 0.029 < 0.8$, which means that, according to the \textit{WISE}-colours, this source should not be classified as an AGN. This seemingly contradicts our findings that this source is a Seyfert galaxy given its position in the BPT-diagram. As \textit{WISE} has a low spatial resolution (namely 6.1~arcsec, in band $W1$ and 6.4~arcsec, in band $W2$), the \textit{WISE}-colour will be a combination of the starlight of the galaxy plus that of the central Seyfert region of the galaxy. Additionally, \citet{LaMassa2019} found that not all AGN are detected by \textit{WISE}, explaining that a non-detection of a known AGN in \textit{WISE} is a possible result of different dust properties, or absence of dust, compared to AGN that are detected by \textit{WISE}, rather than absence of the AGN. We therefore deem our result that NGC~3599 hosts a low-luminosity AGN based on the optical emission line ratios not to be in contradiction of the WISE non-detection. Our conclusion that the nuclear region of NGC~3599 hosts an actively accreting Seyfert-like AGN increases the probability that the observed flare was related to the AGN, although this does not rule out that the XMMJ1115 event was caused by a TDE interacting with the AGN accretion disc \citep{Blanchard2017, Chan2019, Chan2020}.

\begin{table*}
\renewcommand*{\arraystretch}{1.4}
%\begin{center}
\setlength{\tabcolsep}{10pt} % Default value: 6pt
\caption{X-ray luminosities, BH masses and Eddington ratios of host galaxies of known TDEs in low-luminosity AGNs and the host galaxy of XMMJ1115.}
\label{tab:eddingtonratio}
%\hspace*{-2.0cm}
\begin{tabular}{lccccc}
\hline
Transient& Host galaxy & $L_\text{X}$(3--20~keV) & BH mass & log($L/L_\text{Edd}$) \\
name & name & (erg s$^{-1}$) & (M$_\odot$) & & Reference\\
\hline
\textbf{XMMJ1115} & NGC~3599 & 2.25$^{+7.26}_{-1.77}\times$10$^{41}$ * & 2.34$\pm2.27\times$10$^7$ & $-4.11^{+2.15}_{-0.99}$ & a \\
\textbf{ASASSN-14li} & PGC 043234 & 1.32$^{+3.09}_{-0.93}\times$10$^{41}$ & 1.70$^{+2.47}_{-1.02}\times$10$^6$ & $-3.19^{+0.92}_{-0.90}$ & b,c,d \\
\textbf{--} & IC3599 & 6.73$^{+1.96}_{-1.6}\times$10$^{40}$ & 7$\pm5\times$10$^6$ & $-4.11^{+0.65}_{-0.35}$ & e, f, g \\
\textbf{AT2019qiz} & 2MASX~J04463790$-$1013349 & 5.6$^{+20.2}_{4.7}\times$10$^{40}$ * & 1.15$^{+0.85}_{-0.49}\times$10$^6$ & $-3.40^{+0.90}_{-1.0}$ & h\\
\hline 
\end{tabular}
\newline
\textit{Note.} For IC3599 no transient name is listed as this host galaxy has seen multiple flares classified as TDEs since the early 1990s. $L_\text{X}$ marked with * are calculated using the relation between $L_\text{X}$(3--20~keV) and $L_{\text{[\ion{O}{iii}}]}$ from \cite{Heckman2005}, while the other given values are observed $L_\text{X}$ converted to the 3--20~keV band using W3PIMMS$^2$. References: a \citet{Saxton2015}, b \citet{Miller2015nature}, c \citet{VanVelzen2016}, d \citet{Wevers2017}, e \citet{Campana2015}, f \citet{Grupe2001}, g \citet{Grupe2015}, h \citet{Nicholl2020}

%\end{center}
\end{table*}

\begin{figure}
 \includegraphics[width=\columnwidth]{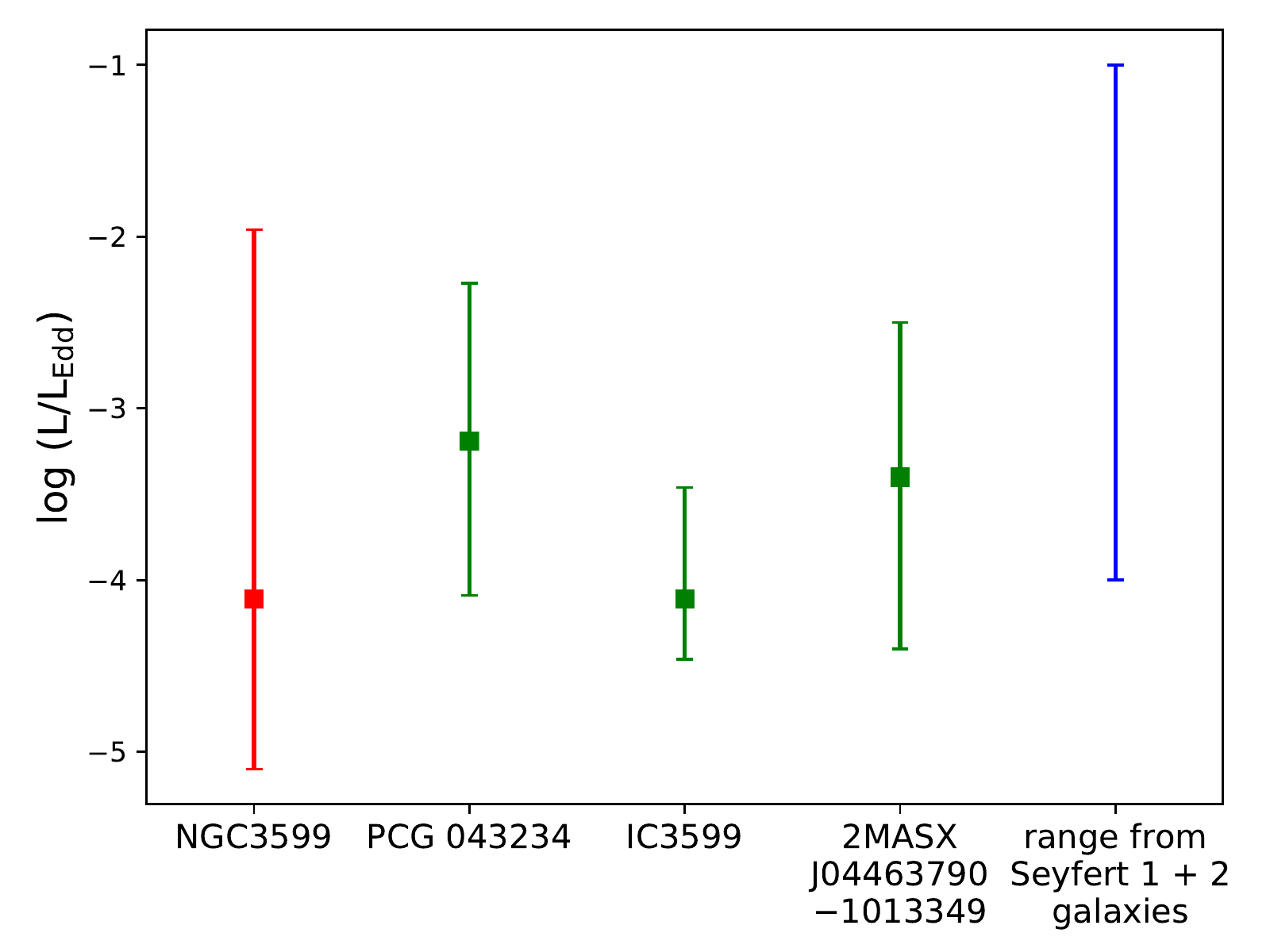}
 \caption{%Plot showing 
 The Eddington ratio for X-ray luminosity of the nucleus three host galaxies of confirmed TDEs in low-luminosity AGNs is shown in green. Our measurement of the Eddington ratio for NGC~3599 based on the spectrum and using the relation from \citet{Heckman2005} to calculate the X-ray luminosity in the 3--20~keV band is shown in red. The blue range shows the Eddington ratios found in Seyfert 1 and Seyfert 2 galaxies by \citet{Singh2011} to represent the range in Eddington ratios for the population of Seyfert galaxies. For the masses, and luminosities used to calculate the Eddington ratio, see Table~\ref{tab:eddingtonratio}.}
 \label{Ledd_comp}
\end{figure}

Assuming NGC~3599 is an AGN and using the empirical relation between $L_\text{X}$(3--20~keV) and $L_{\text{[\ion{O}{iii}}]}$ for AGNs from \citet{Heckman2005}, including the 1$\sigma$ uncertainty in this relation and our 1$\sigma$ uncertainty on the flux measurement, we calculate that $L_\text{X}$(3--20~keV)$= 2.25^{+7.26}_{-1.77} \times 10^{41}$~erg~s$^{-1}$ for NGC~3599. We use this to compute the Eddington ratio of this galaxy in quiescence and compare the value to the observed Eddington ratio of host galaxies of previously confirmed TDEs in low-luminosity AGNs, ASASSN-14li, IC3599 and AT2019qiz (\citealt{Holoien2016, Campana2015, Nicholl2020}, respectively). We list the observed $L_\text{X}$(3--20~keV) values from the literature in Table~\ref{tab:eddingtonratio}, with the X-ray luminosities converted to the 3--20 keV energy band using W3PIMMS\footnote{https://heasarc.gsfc.nasa.gov/cgi-bin/Tools/w3pimms/w3pimms.pl}, as well as our calculated value for NGC~3599. The Eddington ratio calculated for NGC~3599 in quiescence is consistent with Eddington ratios found in other Seyfert galaxies within 1$\sigma$ (see e.g., \citealt{Singh2011}). It is also consistent with the Eddington ratios of the host galaxies of known TDEs in low luminosity AGNs within 1$\sigma$, see Fig~\ref{Ledd_comp}. There we plot the Eddington ratios for the different host galaxies as well as the range of Eddington ratios found by \citet{Singh2011}. The uncertainty in the Eddington ratio for NGC~3599 is dominated by the uncertainty in our flux measurements, in the \citet{Heckman2005} relation and in the black hole mass estimate. Therefore, we favour the conclusion based on the position of the source in the BPT-diagram. We do however note that this comparison of the Eddington ratios also does not exclude a TDE nature of the flare XMMJ1115. 

%From Table~\ref{tab:eddingtonratio} we see that the Eddington ratio of NGC~3599 in quiescence is similar to that of the host galaxies for known TDEs in low luminosity AGNs. This comparison does not provide more insight on the chance of a TDE occuring in this galaxy. We therefore do not alter our conclusion on the nature of the flare XMMJ1115.

%This comparison of the Eddington ratios does not provide additional constraints on the nature of the flare XMMJ1115 and therefore we do not alter our previously stated conclusion.

%and so we conclude this candidate TDE XMMJ1115 could well be due to an AGN flare although a TDE in an AGN disc cannot be ruled out. %still consider this candidate TDE to be more likely due to AGN variability.

%In principle, the different qualifications between the emission line flux and the WISE-colours could also be explained by gas in the nuclear region that is unassociated with the AGN, like in NGC~4993 \citep{Levan2017}, where this nuclear gas around the AGN is not ionized by the AGN itself, but still shows emission line characteristics of LINERs (low-ionization nuclear emission-line regions). This was already shown in \citet{Sarzi2010} and \citet{Singh2013}, who showed that star forming galaxies can appear as LINER galaxies even though they do not show any other signs of AGN activity. However, as NGC~3599 appears as a Seyfert galaxy, not a LINER galaxy, we do not think there is any unassociated gas around the nucleus of this galaxy.

The ratio of the flux of the emission lines in the optical spectrum of the nuclear region of SDSSJ1453 falls in the H~II/starforming region, with the 1$\sigma$ error bar extending into the composite galaxy region (see Fig.~\ref{BPT-PTF09axc}). The \textit{WISE}-colour difference for SDSSJ1453 is $W1-W2 = 0.15 \pm 0.08 < 0.8$, which does not indicate the presence of an AGN in this galaxy. As both the location of the source in a BPT-diagram as well as the \textit{WISE}-colours indicate an inactive galaxy, we conclude that an AGN origin of the observed flare PTF09axc is unlikely, whereas a TDE nature of this transient is consistent with a quiescent galaxy.

\begin{figure}
 \includegraphics[width=\columnwidth]{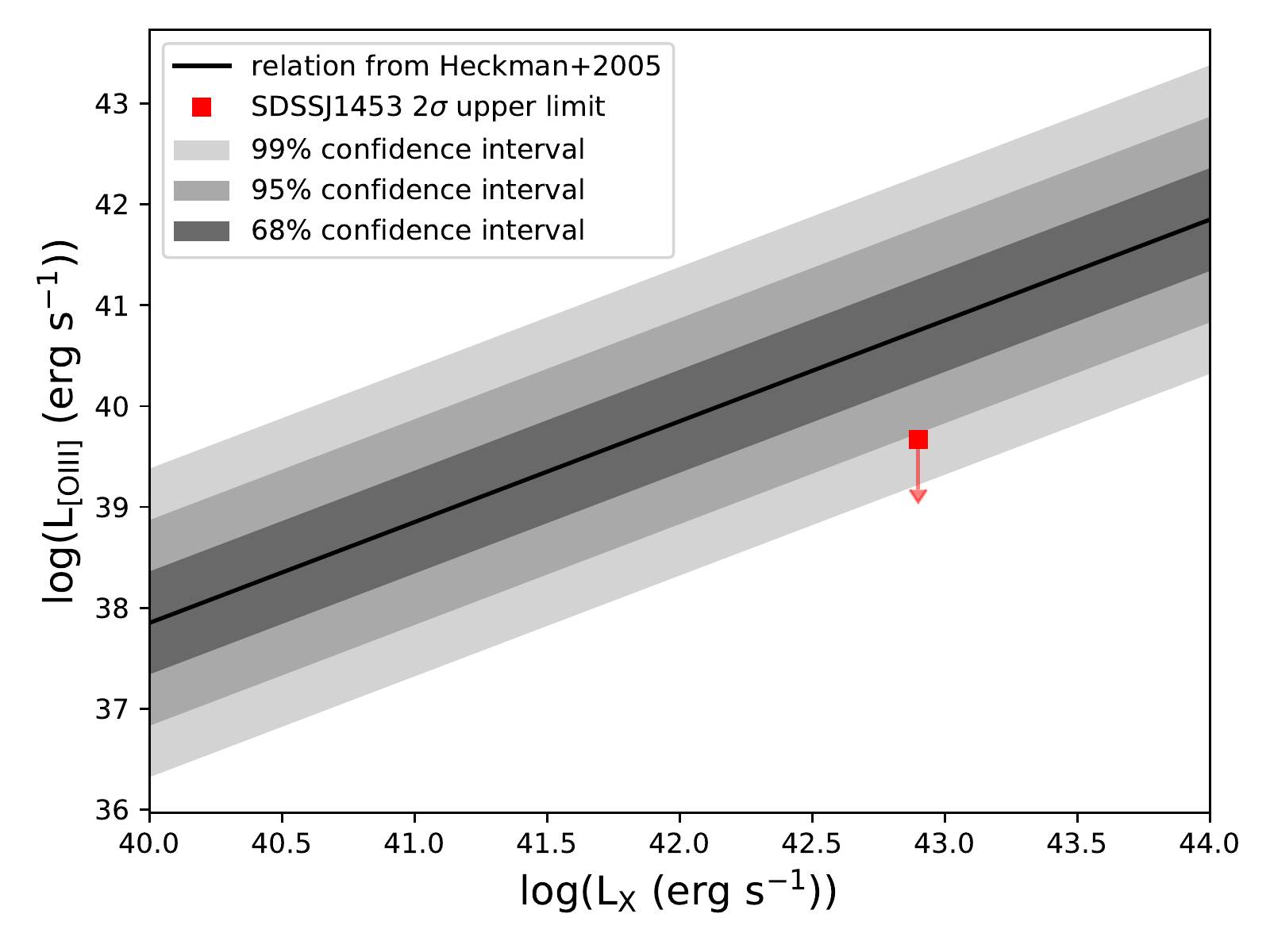}
 \caption{The empirical relation between $L_\text{X}$(3--20~keV) and $L_{\text{[\ion{O}{iii}}]}$ for AGNs from \citet{Heckman2005}  including the 1$\sigma$, 2$\sigma$ and 3$\sigma$ uncertainty regions on the relation in black and increasingly light shades of grey, respectively. In red is our 2$\sigma$ upper limit on $L_{\text{[\ion{O}{iii}}]}$ combined with the measurement for $L_\text{X}$(3--20~keV) from \citet{Jonker2020} to show our upper limit of the host of PTF09axc is not consistent with the relation within 2$\sigma$, but it is within 3$\sigma$. }
 \label{Heckman_rel}
\end{figure}

Using the empirical relation between $L_\text{X}$(3--20~keV) and $L_{\text{[\ion{O}{iii}}]}$ from \citet{Heckman2005} for AGNs (with a scatter of $\sigma = 0.51$~dex or a factor of $\approx3.25$) and $L_\text{X} = 8 \times 10^{42}$~erg s$^{-1}$ from \citet{Jonker2020}, the predicted AGN luminosity from the [\ion{O}{iii}] line would be $L_{\text{[\ion{O}{iii}}]} \approx 5.7 \times 10^{40}$~erg~s$^{-1}$. Using our 2$\sigma$ upper limit for the flux in the emission line of $1.39 \times 10^{-16}$~erg~cm$^{-2}$~s$^{-1}$ and d$_L = 536.1$~Mpc we calculate $L_{\text{[\ion{O}{iii}}]} = 4.78 \times 10^{39}$~erg~s$^{-1}$, which is a factor 11.9 lower than expected for an AGN-powered emission line (see Figure~\ref{Heckman_rel} where we show the relation and our upper limit on $L_{\text{[\ion{O}{iii}}]}$). While our observed upper limit for the $L_{\text{[\ion{O}{iii}}]}$ is too low compared to the luminosity predicted by the correlation from \citet{Heckman2005}, it is consistent if we take the uncertainty on this correlation into account within 3$\sigma$. This means we cannot exclude that the PTF09axc host SDSSJ1453 is an AGN, based on the $L_X~\propto$~[\ion{O}{iii}] relation from \citet{Heckman2005}. However, the combined evidence provided by the upper limit on $L_{\text{[\ion{O}{iii}}]}$, from the position of this source on the BPT-diagram, and the \textit{WISE}-colour difference, we conclude that this galaxy is most likely quiescent and the transient PTF09axc is most likely a TDE.

%However, if we take this uncertainty into account, then our observed upper limit the observed 95\% confidence upper limit to our $L_{\text{[\ion{O}{iii}}]}$ is consistent with the expected $L_{\text{[\ion{O}{iii}}]}$ within $2\sigma$. However, as this is only an upper limit, it is not explicit proof that the empirical relation between $L_\text{X}$(3-20~keV) and $L_{\text{[\ion{O}{iii}}]}$ holds for this source. 

It should be noted that in all low- to medium-resolution spectroscopic ground based observations  the [\ion{O}{iii}]~$\lambda5007$ emission line in SDSSJ1453 is redshifted to fall close to the [\ion{O}{i}]~$\lambda 5577$ terrestrial sky line, although in previous work (e.g., \citealt{Arcavi2014}) there is no mention of this. Their $L_{\text{[\ion{O}{iii}}]}$ is consistent with our upper limit and not consistent with the empirical relation between $L_\text{X}$(3--20~keV) and $L_{\text{[\ion{O}{iii}}]}$ and therefore supports our conclusion about PTF09axc.

\section*{Acknowledgements}

D.M.S. acknowledges support from the ERC under the European Union’s Horizon 2020 research and innovation programme (grant agreement No. 715051; Spiders). This work is part of the research programme Athena with project number 184.034.002, which is financed by the Dutch Research Council (NWO). We thank Tom Marsh for the use of {\sc molly}. This work uses {\sc python} packages {\sc numpy, lmfit, matplotlib, sys, astropy}.

%%%%%%%%%%%%%%%%%%%%%%%%%%%%%%%%%%%%%%%%%%%%%%%%%%
\section*{Data Availability}

%A package for the reproduction of figures and is availabe on Zenodo (

All data will be made available in a reproduction package uploaded
to Zenodo.
 
%The inclusion of a Data Availability Statement is a requirement for articles published in MNRAS. Data Availability Statements provide a standardised format for readers to understand the availability of data underlying the research results described in the article. The statement may refer to original data generated in the course of the study or to third-party data analysed in the article. The statement should describe and provide means of access, where possible, by linking to the data or providing the required accession numbers for the relevant databases or DOIs.



\bibliographystyle{mnras}
\bibliography{references} % if your bibtex file is called example.bib

%\begin{thebibliography}{99}
%\end{thebibliography}

%%%%%%%%%%%%%%%%%%%%%%%%%%%%%%%%%%%%%%%%%%%%%%%%%%

%%%%%%%%%%%%%%%%% APPENDICES %%%%%%%%%%%%%%%%%%%%%

\appendix

\section{Additional Material}

\begin{figure*}
 \centering
 \begin{subfigure}{\columnwidth}
  \includegraphics[width=\columnwidth]{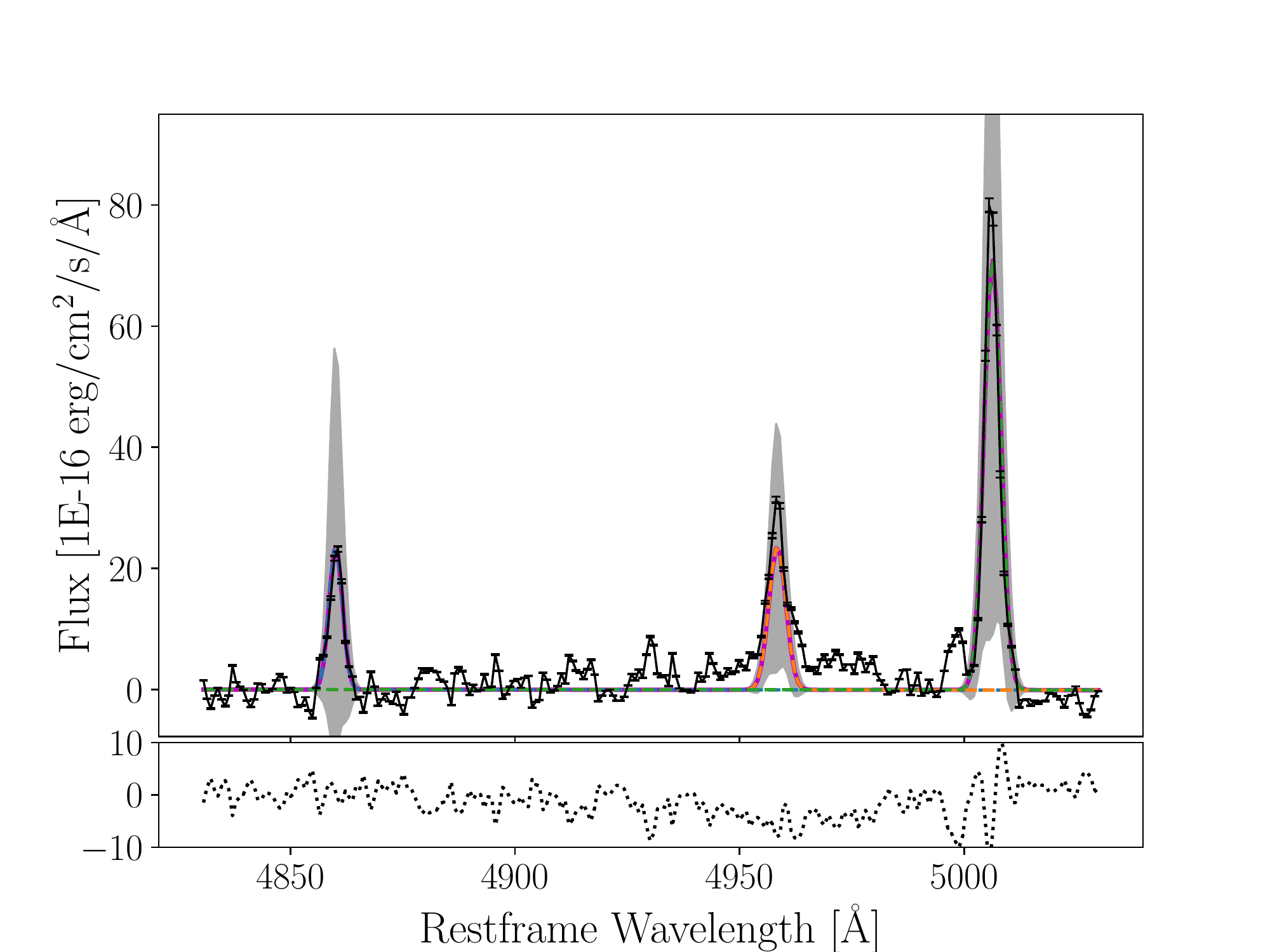}
  \caption{}
 \end{subfigure}
 \begin{subfigure}{\columnwidth}
  \includegraphics[width=\columnwidth]{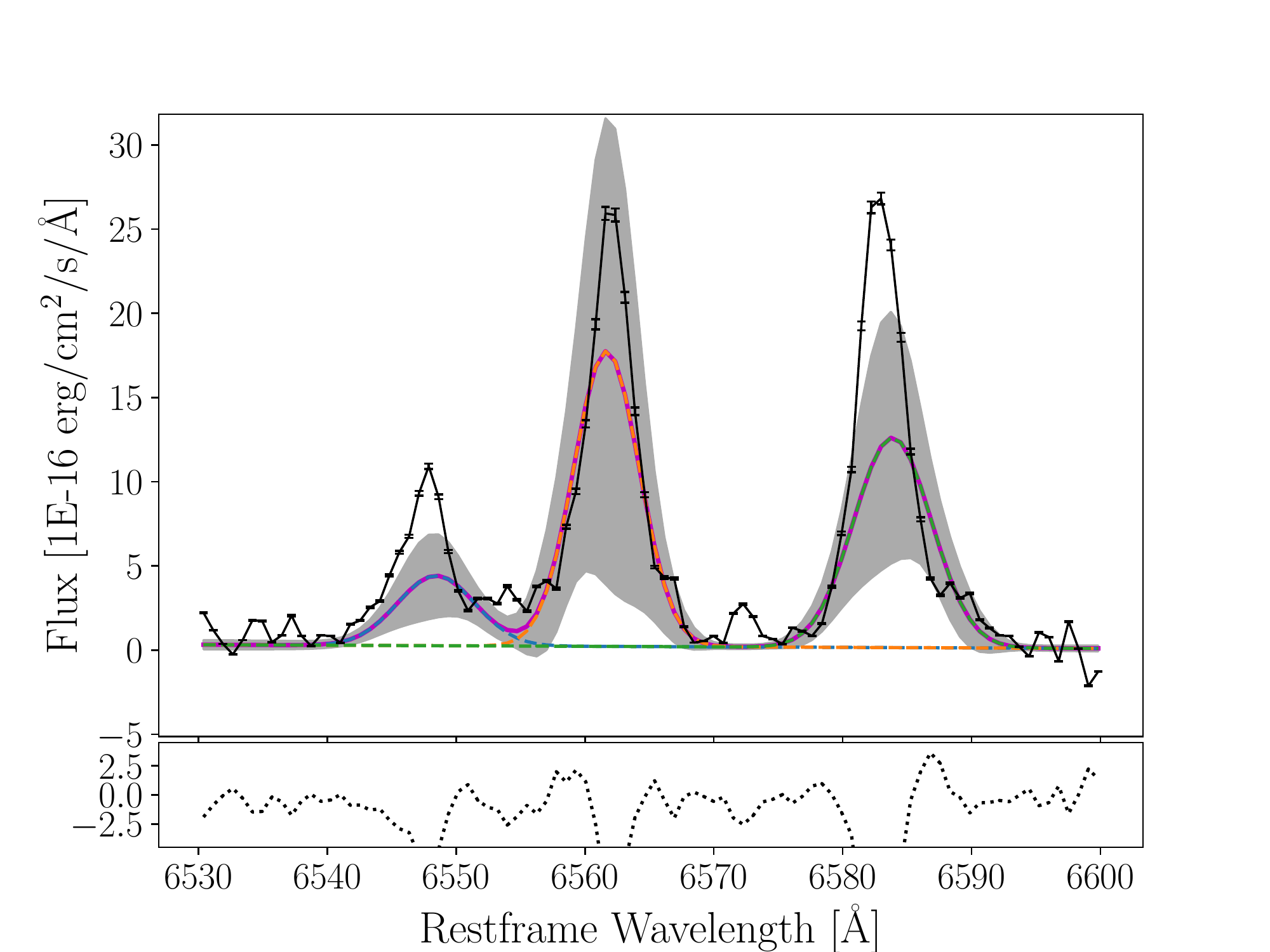}
  \caption{}
 \end{subfigure}
 \begin{subfigure}{\columnwidth}
  \includegraphics[width=\columnwidth]{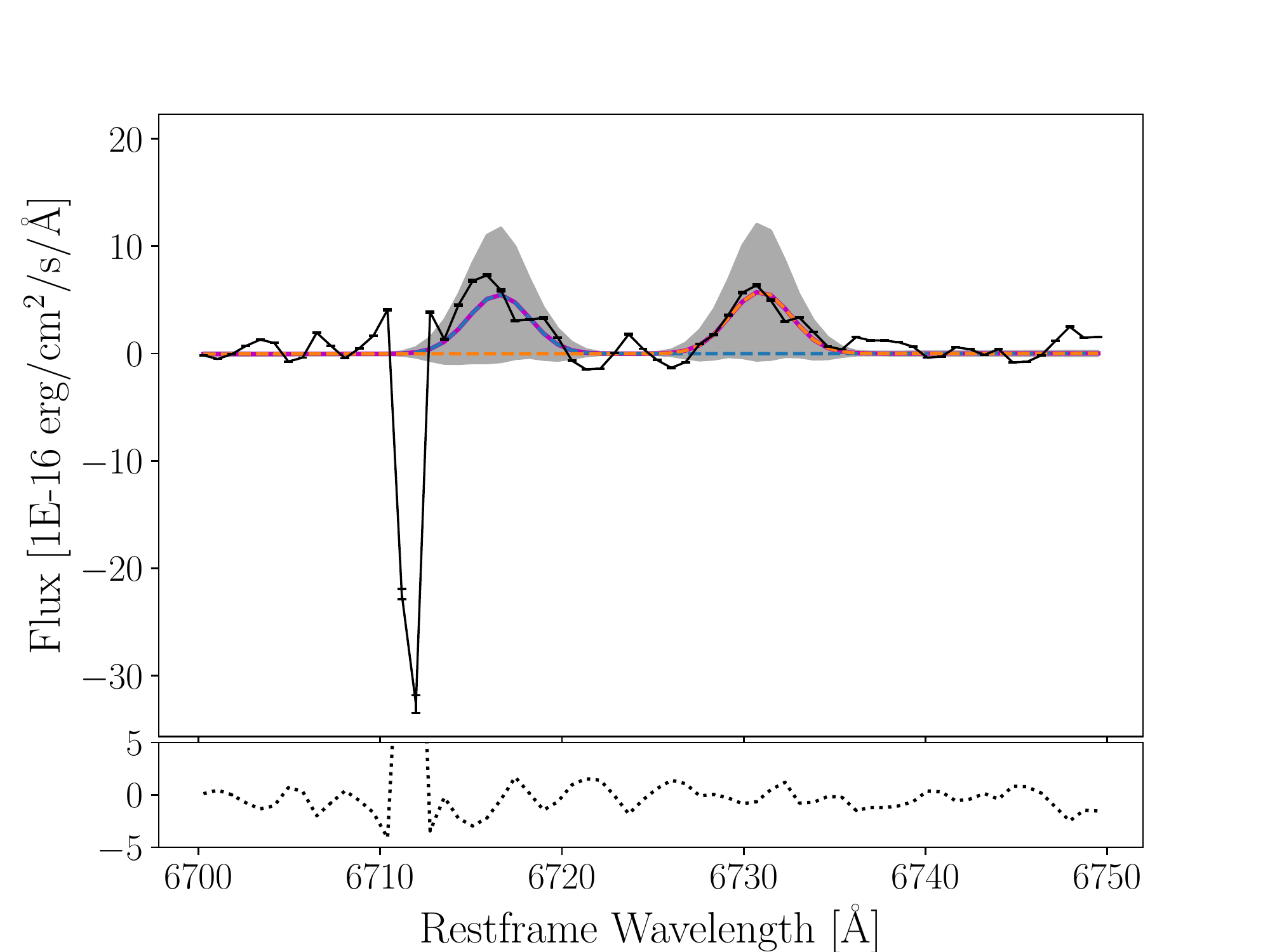}
  \caption{}
 \end{subfigure}
 \caption{Fit to the emission lines of NGC~3599 after subtracting the starlight component of the host galaxy spectrum. The data are shown in black, the total best fit is shown in magenta, the individual components have varying colours. (a) shows H~$\beta$ (blue) and [\ion{O}{iii}]~$\lambda4959$ (orange) and $\lambda5007$ (green), (b) shows H~$\alpha$ (orange) and [\ion{N}{ii}]~$\lambda6548$ (blue) and $\lambda6584$ (green) and (c) shows [\ion{S}{ii}]~$\lambda6717$ (blue) and $\lambda6731$ (orange). Error bars to the spectrum are indicated in all frames in black, the grey shaded area corresponds to the 3$\sigma$ region of the best fit. }
 \label{NGC3599_lines}
\end{figure*}

\begin{figure*}
 \centering
 \begin{subfigure}{\columnwidth}
  \includegraphics[width=\columnwidth]{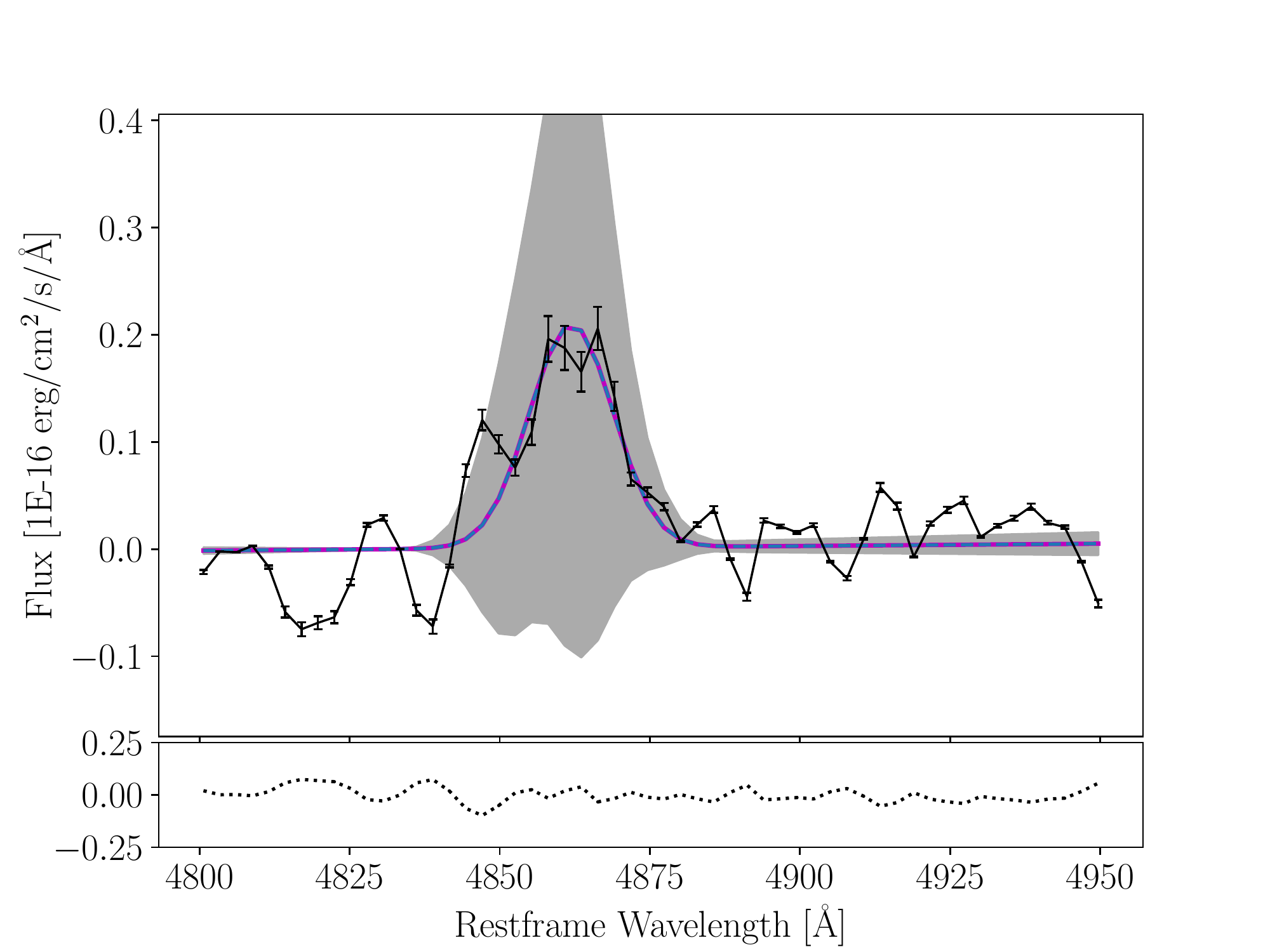}
  \caption{}
 \end{subfigure}
 \begin{subfigure}{\columnwidth}
  \includegraphics[width=\columnwidth]{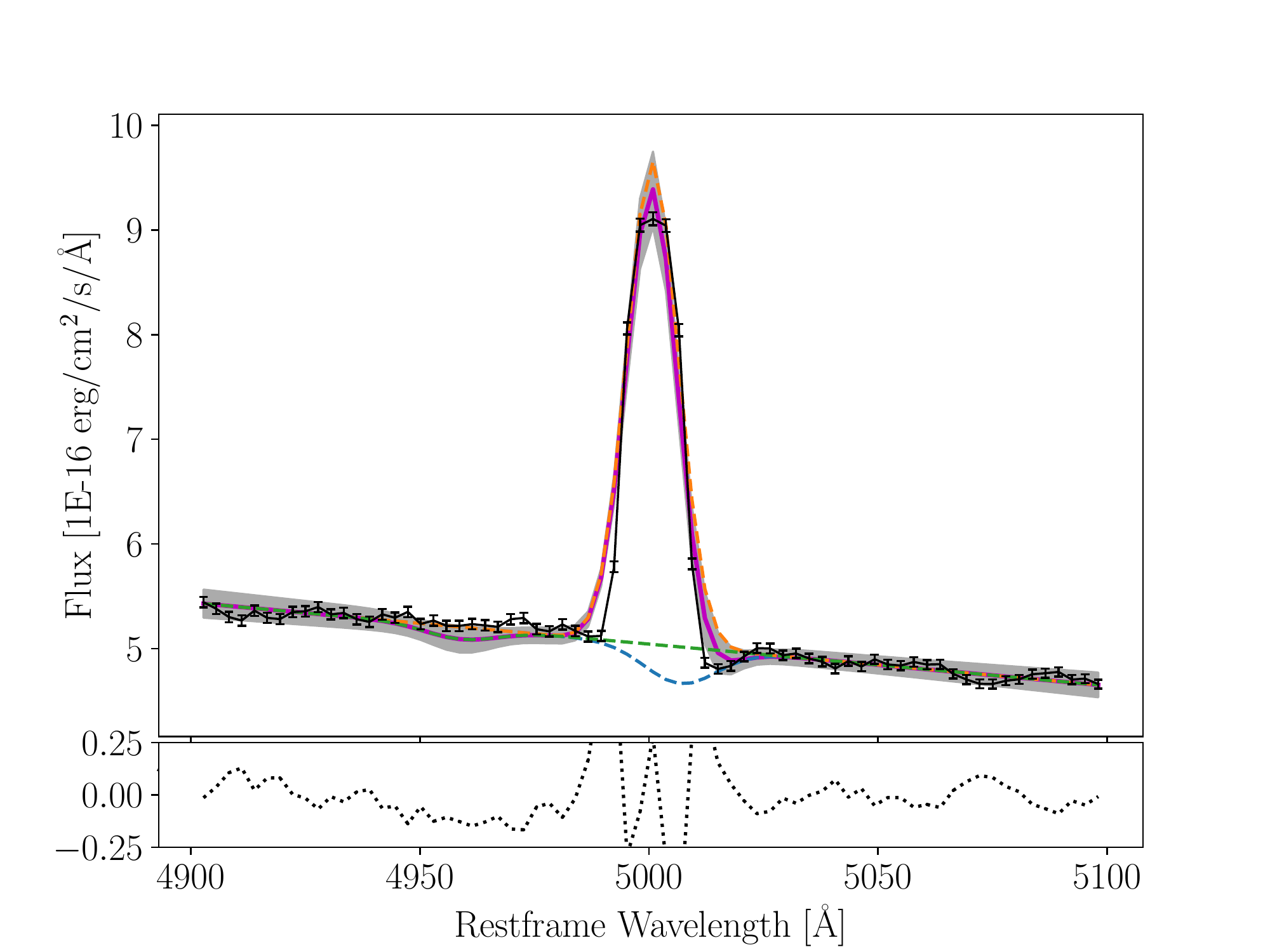}
  \caption{}
 \end{subfigure}
 \begin{subfigure}{\columnwidth}
  \includegraphics[width=\columnwidth]{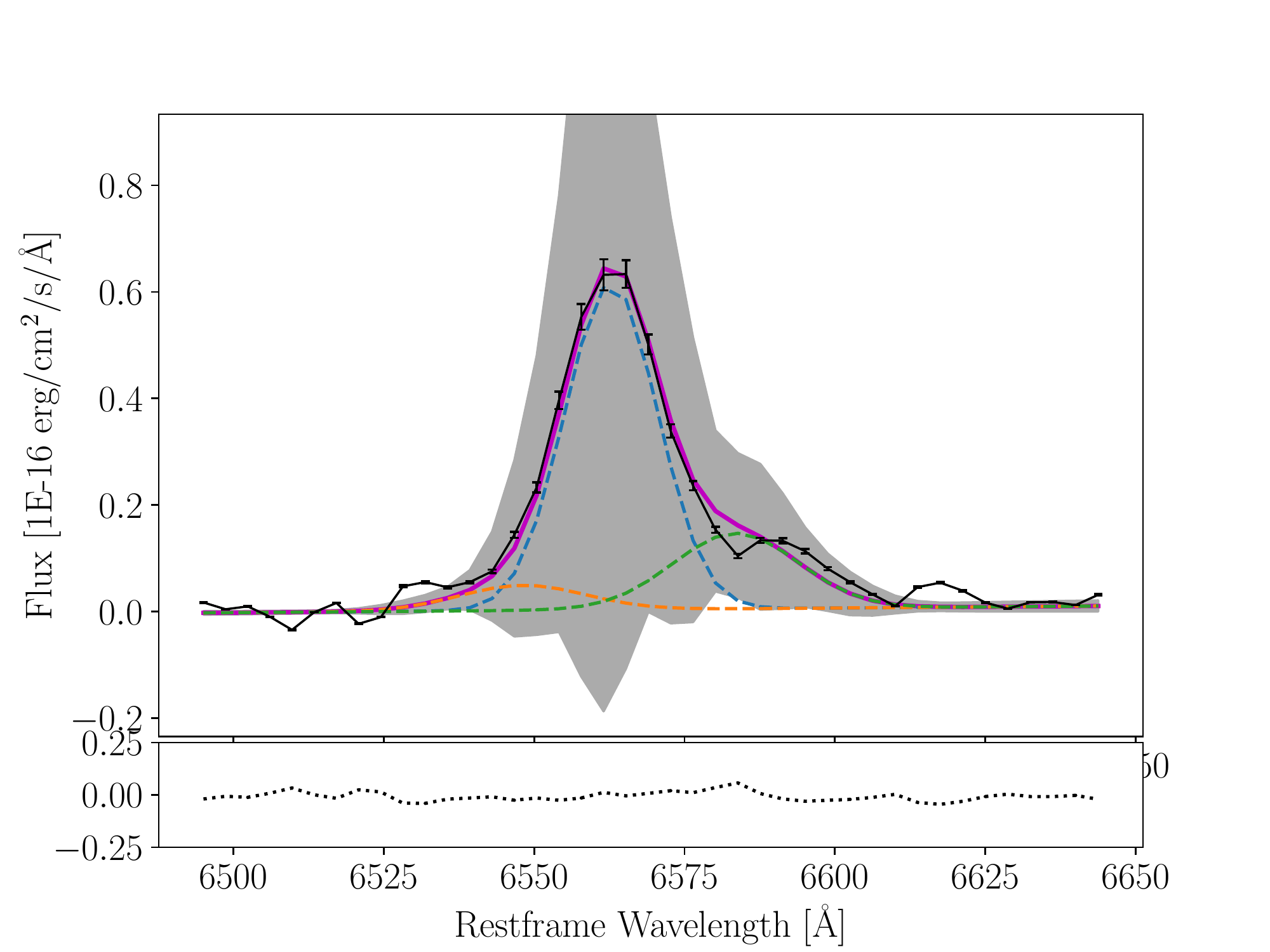}
  \caption{}
 \end{subfigure}
 \caption{Fit of the emission lines of the hosty galaxy of PTF09axc with the starlight component subtracted for (a) and (c) and without the background or the starlight component subtracted for (b). The data are shown in black, the total best fit is shown in magenta, the individual components have varying colours. (a) shows H~$\beta$ (blue), (b) shows \ion{O}{i}~$\lambda5577$ terrestrial skyline (orange) and [\ion{O}{iii}]~$\lambda4959, 5007$ line (green and blue respectively) and (c) shows H~$\alpha$ (blue) and [\ion{N}{ii}]~$\lambda6548, 6584$ (orange and green respectively). Error bars to the spectrum are indicated in all frames in black, the grey shaded area corresponds to the 3$\sigma$ region of the best fit.}
 \label{PTF09axc_lines}
\end{figure*}

\begin{landscape}
\begin{table}
%\begin{rotatetable*}
%\begin{center}
\setlength{\tabcolsep}{10pt} % Default value: 6pt
\caption{Results of line fitting of the most prominent emission lines in the host galaxy nuclear spectra of TDE candidates XMMJ1115 and PTF09axc.}
\label{tab:lines}

%\hspace*{-4.5cm}
\begin{tabular}{lcccccc}
\hline
& \multicolumn{3}{c}{[\ion{O}{iii}]~$\lambda4959$} & \multicolumn{3}{c}{[\ion{O}{iii}]~$\lambda5007$}\\
source & WL [$\textrm{\AA}$] & FWHM [km~s$^{-1}$] & flux [1E-16~erg~cm$^{-2}$~s$^{-1}$] & WL [$\textrm{\AA}$] & FWHM [km~s$^{-1}$] & flux [1E-16~erg~cm$^{-2}$~s$^{-1}$] \\
\hline
\textbf{XMMJ1115}& 4958.4$\pm$0.3$^\dagger$&270.0$\pm$27.8$^*$&112.45$\pm$34.36 &5006.3$\pm$0.3$^\dagger$&267.5$\pm$27.5$^*$&337.35$\pm$103.07 \\
\textbf{PTF09axc}& \textit{4958.94$\pm$0.00}$^\dagger$ & \textit{774.17$\pm$0.00}$^*$ & 0.46 &\textit{5006.84$\pm$0.0}$^\dagger$&\textit{766.77$\pm$0.0}$^*$ & 1.39 \\
\hline
\hline
& \multicolumn{3}{c}{H~$\beta$}& \multicolumn{3}{c}{H~$\alpha$}\\
source & WL [$\textrm{\AA}$] & FWHM [km~s$^{-1}$] & flux [1E-16~erg~cm$^{-2}$~s$^{-1}$] & WL [$\textrm{\AA}$] & FWHM [km~s$^{-1}$] & flux [1E-16~erg~cm$^{-2}$~s$^{-1}$] \\
\hline
\textbf{XMMJ1115}&4860.1$\pm$0.4&215.9$\pm$30.2&87.80$\pm$43.17 & 6561.7$\pm$0.3&272.8$\pm$24.6&111.48$\pm$30.31 \\
\textbf{PTF09axc}&4861.9$\pm$2.1&605.3$\pm$200.1&2.18$\pm$1.24& 6562.8$\pm$1.9$^\ddagger$&569.9$\pm$219.1&8.14$\pm$4.57 \\
\hline
\hline
 &\multicolumn{3}{c}{[\ion{N}{ii}]~$\lambda6548$}& \multicolumn{3}{c}{[\ion{N}{ii}]~$\lambda6584$}\\
source & WL [$\textrm{\AA}$] & FWHM [km~s$^{-1}$] & flux [1E-16~erg~cm$^{-2}$~s$^{-1}$] & WL [$\textrm{\AA}$] & FWHM [km~s$^{-1}$] & flux [1E-16~erg~cm$^{-2}$~s$^{-1}$] \\
\hline
\textbf{XMMJ1115}& 6547.5$\pm$0.3$^\ddagger$&324.0$\pm$25.1$^{**}$&31.32$\pm$6.49&6583.9 $\pm$0.3$^\ddagger$&322.2$\pm$25.0$^{**}$&93.95$\pm$19.48 \\
\textbf{PTF09axc}& 6548.1$\pm$0.0$^\ddagger$&950.2$\pm$197.8$^{**}$&1.04$\pm$0.47& 6583.5$\pm$1.9$^\ddagger$ &945.1$\pm$196.8$^{**}$&3.13$\pm$1.41 \\
\hline
\hline
& \multicolumn{3}{c}{[\ion{S}{ii}]~$\lambda6717$} & \multicolumn{3}{c}{[\ion{S}{ii}]~$\lambda6731$} \\
source & WL [$\textrm{\AA}$] & FWHM [km~s$^{-1}$] & flux [1E-16~erg~cm$^{-2}$~s$^{-1}$] & WL [$\textrm{\AA}$] & FWHM [km~s$^{-1}$] & flux [1E-16~erg~cm$^{-2}$~s$^{-1}$] \\
\hline
\textbf{XMMJ1115}  & 6716.5$\pm$0.3$^+$&177.1$\pm$24.3$^-$&23.28$\pm$9.08 & 6730.9$\pm$0.3$^+$&176.8$\pm$24.2$^-$&24.37$\pm$9.28\\
\textbf{PTF09axc} & $\cdots$ &$\cdots$ &$\cdots$ & $\cdots$&$\cdots$&$\cdots$\\
\hline
\end{tabular}
\newline
\textit{Note.} With $\cdots$ we indicate that this line could not be fitted to the data. Different markers indicate quantities that were tied to the same value (FWHM) or a set separation\\
(WL) for each of the sources. Numbers in italics were forced to the mentioned value to obtain a 2$\sigma$ upper limit (UL) on the flux of that emission line. Flux measurements\\ without error are 2$\sigma$ upper limits. 

%\end{center}
%\end{rotatetable*}
\end{table}
\end{landscape}

%%%%%%%%%%%%%%%%%%%%%%%%%%%%%%%%%%%%%%%%%%%%%%%%%%

% Don't change these lines
\bsp	% typesetting comment
\label{lastpage}
\end{document}